\documentstyle[ 12pt]{article}

\begin{document}

\centerline { \bf MAXIMAL-ACCELERATION PHASE SPACE RELATIVITY}

\centerline{\bf  FROM CLIFFORD ALGEBRAS }

\bigskip

\centerline { Carlos Castro }

\centerline{ Center for Theoretical Studies of Physical Systems, }

\centerline{ Clark Atlanta University, Atlanta, GA. 30314 }

\centerline { August, 2002 }

\bigskip

\centerline{\bf  Abstract}

\noindent
We present a new physical model that links the maximum speed of light with
the minimal Planck scale into a maximal-acceleration Relativity principle in
the spacetime tangent bundle and in phase spaces (cotangent bundle). Crucial
in order to establish this link is the use of Clifford algebras in phase
spaces. The maximal proper-acceleration bound is  $a = c^2/ \Lambda$  in 
full
agreement with the old predictions of Caianiello, the Finslerian geometry
point of view of Brandt and more recent results in the literature. We
present the reasons why an Extended Scale Relativity based on Clifford 
spaces
is physically more appealing than those based on kappa-deformed Poincare
algebras and the inhomogeneous quantum groups operating in quantum Minkowski
spacetimes. The main reason being that the Planck scale should not be taken
as a deformation parameter to construct quantum algebras but should exist
already as the minimum scale in Clifford spaces.

\bigskip

\centerline {\bf I .  Introduction }

\bigskip

Relativity in C-spaces (Clifford manifolds) [1] is a very natural extension
of Einstein's relativity and Nottale's scale relativity [2] where the
impassible speed of light and the minimum Planck scale are the two universal
invariants.  An event in C-space is represented by a {\em polyvector}, or
Clifford-aggregate of lines, areas, volumes. which  bear a one-to-one
correspondence to the holographic shadows/projections (onto the embedding
spacetime coordinate planes) of a nested family of $p$-loops (closed $p$-
branes of spherical topology) of various dimensionalities: $p = 0$ 
represents
a point; $p = 1$ a closed string, $p = 2$ a closed membrane, etc.

The invariant ``line"  element associated with a polyparticle is:

$$ d\Sigma ^2 = dX.dX =   d \Omega) ^2 + \Lambda ^{2D-2} ( dx_{\mu}dx^
{\mu} ) +
\Lambda ^{ 2D -4 } ( d x_{\mu\nu} ) ( d x^{\mu\nu} ) + ...\eqno(1.1) $$
the Planck scale appears as a natural quantity in order to match units and 
combine p-branes ( p-loops ) of different dimensions. The fact that the 
Planck scale is a minimum was based on the real-valued interval $ dX $ when 
$ dX.dX  > 0 $. The analog of photons in C-space are $tensionless$ branes : $ 
dX.dX = 0 $. Scales smaller than
$\Lambda$ yield  " tachyonic  " intervals $ dX. dX  < 0 $ [1]. Due
to the matrix representation of the gamma matrices and the cyclic trace
property, it can be easily seen why the line element is invariant under the 
C-
space Lorentz $group$  transformations:
$$Trace ~ X'^{2 } = Trace~ [ R X^2 R^{ - 1 } ] = Trace~ [ R R^{ -1 } X^2 ] =
Trace~X^2 ~, \eqno( 1.2) $$
where a finite polydimensional rotation that reshuffles dimensions is
characterized by the C-space ``rotation" matrix:

$$  R = \exp [ i (\theta I + \theta ^{\mu} \gamma _{\mu} + \theta^{\mu\nu}
\gamma _{\mu \nu} + ...)] .  \eqno(1.3) $$
The parameters $\theta, \theta ^{\mu}, \theta ^{\mu\nu}$ are the C-space
extension of the Lorentz boost parameters and for this reason the naïve
Lorentz transformations of spacetime are modified to be:

$$ x'^{\mu} = L^{\mu}_{\nu} ~ [ \theta , \theta ^\mu, \theta^{\mu\nu} , ...]
x^{\nu}~. \eqno (1.4 ) $$

Due to the presence of all the C-space boost parameters in the modified
Lorentz matrix
$L^{\mu}_{\nu}$ one induces an effective Lorentz transformation with an
$effective$ boost parameter of the
form  $\xi_{eff} = z ( \xi )$ discussed in [12] in order to ensure that the
minimal Planck scale is never surpassed at infinite momentum.
For details we refer to [1].

It was argued in [1] that the extended Relativity principle in C-space may
contain the clues to unravel the physical foundations of string and M-theory
since the dynamics in C-spaces encompass in one stroke the dynamics of all 
p-
branes of various dimensionalities. In particular, how to formulate a master
action that encodes the collective dynamics of all extended objects.

For further details about these issues we refer to  [1,3,16] and all the
references therein. Like the derivation of the minimal length/time
string/brane uncertainty relations;  the logarithmic corrections to the 
black-
hole area-entropy relation;  the origins of a higher derivative gravity with
torsion; the construction of the p-brane propagator;  the role of
supersymmetry;  the emergence of two times; the reason behind a running
Planck constant and the variable fine structure constant; the way to
correctly pose the cosmological constant problem as well as other results.

In this letter, we will present another physical       that links the 
maximum
speed of light and the minimal Planck scale into a maximal-acceleration
principle in the spacetime tangent bundle, and consequently, in the phase
space (cotangent bundle). Crucial in order to establish this link is the use
of Clifford  algebras in phase spaces. The maximal proper acceleration bound
is  $a = c^2/ \Lambda$  in full agreement with [4] and the Finslerian
geometry point of view in [6].

In section {\bf II}, we show how to derive the Nesterenko action [5]
associated with a sub-maximally accelerated particle in spacetime directly
from phase-space Clifford algebras and present a  full-fledged C-phase-space
generalization of the Nesterenko action associated with the multi-symplectic
geometry of a polyparticle.

In {\bf III}, we present a series of reasons why we believe C-space
Relativity is more physically appealing  than all the others proposals based
on kappa-deformed Poincare algebras and other quantum algebras [10-13,17]. 
We
also argue why the theories based on kappa-deformed Poincare algebras may in
fact be related to a Moyal star-product deformation of a classical Lorentz
algebra whose deformation parameter is precisely the Planck scale $\Lambda =
1/ \kappa$.

 \bigskip

\centerline { \bf  II.  Maximal-Acceleration from Clifford algebras }

 \bigskip

We will follow closely the procedure described in the book [3] to construct
the phase space Clifford algebra. For simplicity we shall begin with a two--
dimensional phase space, with one coordinate and one momentum variable and
afterwards we will generalize the construction  to higher dimensions.

Let $ e_{p}   e_{q} $ be the Clifford basis elements in a two--dimensional
phase space obeying the following relations:

$$  e_p . e_q \equiv  { 1 \over 2 } ( e_q e_p + e_p e_q ) = 0. ~~~ e_p . e_p
= e_q . e_q = 1 . \eqno(2.1) $$

The Clifford product of  $ e_p, e_q $ is by definition the sum of the scalar
product and wedge product furnishing the unit $bivector$:

 $$ e_p e_q \equiv e_p . e_q +  e_p \wedge e_q  = e_p \wedge e_q =  j  . 
~~~
j^ 2 =  e_p e_q  e_p e_q = - 1 . \eqno (2.2 ) $$
due to the fact that $ e_p, e_q $ anticommute, eq.~( 2.1).

In this fashion, using Clifford algebras one can justify the origins of
complex numbers without introducing them ad-hoc. The imaginary unit $ j $ is
$ e_p e_q $.  For example, a Clifford vector in phase space can be expanded,
setting aside for the moment the issue of units, as:

$$  Q = q e_q  + p e_p . ~~~ Q e_q = q + p e_p e_ q = q + j p = z. ~~~
e_q Q = q + p e_q e_p = q - j p = z^*~, \eqno ( 2.3 ) $$
which reminds us of the creation/annhilation operators used in the harmonic
oscillator case and in coherent states.

The analog of the action for a massive particle is obtained by taking the
scalar product:
$$ dQ . dQ = (dq)^2 + ( dp)^2  \Rightarrow S = m \int  \sqrt { dQ. dQ } =
m \int  \sqrt {( dq)^2 + (dp)^2 } . \eqno ( 2.4 ) $$

One may insert now the appropriate length and mass parameters in order to
have consistent units:

$$ S = m \int  \sqrt {   dq)^2 + ( {  \Lambda \over m  } )^2  (dp)^2 }.
\eqno (2.5 ) $$
where we have introduced  the Planck scale $\Lambda$  and the mass $m$ of 
the
particle to have consistent units, $\hbar = c = 1$.
The reason will become clear below.

Extending this two-dimensional action to a higher $2n$-dimensional phase
space requires to have
$ e_{p_{\mu }} ,  e_{q_{\mu}} $ for the Clifford basis where $\mu = 1, 2,
3...n$. The action in this $2n$-dimensional phase space is:
$$ S = m \int  \sqrt {   dq^{\mu} dq_{\mu} ) + ({\Lambda \over m  } )^2  
(dp^
{\mu} dp_{\mu})} =
m \int   \tau \sqrt {1  + ( {  \Lambda \over m  } )^2  (dp^{\mu}/ d \tau )
(  dp_{\mu}/ d \tau   }
\eqno (2.6 ) $$
in units of $c = 1$, one has the usual infinitesimal proper time
displacement  $d \tau ^2 = dq^{\mu}  dq_{\mu}$.

One can easily recognize that this action (2.6), up to a numerical factor of
$ m/a$,  is nothing but the action for a sub-maximally accelerated particle
given by
Nesterenko  [5].  It is sufficient to rewrite: $dp^\mu / d \tau  = m d^2
x^\mu / d \tau ^2 $  to get from eq.~(2.6):
$$ S = m \int   \tau \sqrt {  1  + \Lambda^2  (d^2 x^ \mu/ d \tau^2   (
d^2 x_\mu/ d \tau^2) } . \eqno ( 2.7) $$

Using the postulate that the maximal-proper acceleration is given in a
consistent manner with the minimal length principle (in units of $c = 1$):

$$ a = c^2 / \Lambda  = 1/\Lambda \Rightarrow
S = m \int   \tau \sqrt {  1  +   (  { 1 \over a }  )^2  (d^2 x^ \mu/ d
\tau^2   (  d^2 x_\mu/ d \tau^2   ) }~, \eqno ( 2. $$
which is exactly the action of [5], up to a numerical factor of $ m/a $,
when  the metric signature is $ ( +, -, - , - ) $.

The proper acceleration is $orthogonal$ to the proper velocity as a result 
of
differentiating the timelike proper velocity  squared:

$$ V^2 = {dx^{\mu} \over d \tau } { d x_{\mu} \over d \tau } = 1 =   V^\mu
V_\mu  > 0 \Rightarrow  { d V^{\mu} \over d \tau } V_\mu =  { d^2 x^{\mu}
\over d \tau ^2 }
V_\mu =  0~, \eqno(2.9)  $$
which means that if the proper velocity is timelike the proper acceleration
is spacelike so that:

 $$  g^2 ( \tau ) \equiv -  (d^2 x^ \mu/ d \tau^2   (  d^2 x_\mu/ d 
\tau^2)
 > 0  \Rightarrow
S = m \int   \tau \sqrt {  1  -    { g^2  \over a^2  }     } \equiv  m \int
d \omega ~, \eqno(2.10 ) $$
where we have defined:

$$   \omega  \equiv \sqrt {  1  -    { g^2  \over a^2  }     } d \tau . 
\eqno
(2.11) $$
The dynamics of a submaximally accelerated particle in Minkowski spacetime
can be reinterpreted as that of a particle moving in the spacetime $tangent-
bundle$ background whose $Finslerian$-like metric is:

$$  d\omega^2 = g_{\mu \nu} ( x^\mu, dx^\mu ) dx^\mu dx^\nu  =
(d \tau)^2    1  -    { g^2  \over a^2  }     )                . 
\eqno(2.12)
$$

For uniformly accelerated motion, $ g ( \tau ) = g = constant$ the factor:

$$ { 1 \over \sqrt {  1  -  { g^2  \over a^2  }  }   }  \eqno (2.13) $$
plays a similar role as the standard Lorentz time dilation factor in
Minkowski spacetime.

The action is real valued if, and only if, $ g^2  < a^2 $ in the same way
that the action in Minkowski spacetime is real valued if, and only if, $v^2 
<
c^2$. This explains why the particle dynamics has a bound on proper-
accelerations. Hence, for the particular case of a $uniformly$ accelerated
particle whose trajectory in Minkowski spacetime is a hyperbola, one has an
Extended  Relativity of $uniformly$ accelerated observers whose proper-
acceleration have an upper bound given by  $c^2/ \Lambda$. Rigorously
speaking, the spacetime trajectory is obtained by a canonical projection of
the spacetime tangent bundle onto spacetime. The invariant time, under the
pseudo-complex extension of the Lorentz group [8], measured in the spacetime
tangent bundle is no longer the same as $\tau$,  but instead, it is given by
$\omega ( \tau )$.

This is similar to what happens in C-spaces, the truly invariant evolution
parameter is not
$\tau$ nor $\Omega$, the Stuckelberg parameter [3], but it is $\Sigma$  
which
is the world interval in C-space and that has units of $length ^ D$. The
$group$ of C-space Lorentz transformations preserve the norms of the
Polyvectors and these have units of hypervolumes; hence C-space Lorentz
transformations are volume-preserving.

Another approach to obtain the action for a sub-maximally accelerated
particle was given by [8] based on a pseudo-complexification of Minkowski
spacetime and the Lorentz group that describes the physics of the spacetime
tangent bundle. This picture is not very different form the Finslerian
spacetime tangent bundle point of view of Brandt [6].  The invariant group 
is
given by a pseudo-complex extension of the Lorentz group acting on the
extended coordinates $ X = a x^\mu + I v^\mu $ with $ I^2 =  1 $ (pseudo-
imaginary unit) where both position and velocities are unified on equal
footing. The invariant line interval is
$a^2 d \omega^2 = (dX)^2$.

A C-phase-space generalization of these actions (for sub-maximally
accelerated particles, maximum tidal forces) follows very naturally by using
polyvectors:
$$ Y =  q^\mu e_{ q_\mu  } +   q^ { \mu \nu } e_{ q_\mu  } \wedge e_{
q_\nu  } +
q^ { \mu \nu\rho } e_{ q_\mu  } \wedge e_{ q_\nu  } \wedge e_{q_\rho} + 
....$$
$$ + p^\mu e_{ p_\mu  } +   p^ { \mu \nu } e_{ p_\mu  } \wedge e_{ p_\nu  }
+ ...~,\eqno (2.14) $$
where one has to insert suitable powers of $\Lambda$ and $m$ in the 
expansion
to match units.

The C-phase-space action reads then:

$$S \sim \int \sqrt { dY . dY } = \int \sqrt {   dq^\mu dq_\mu +   q^{\mu
\nu } dq_{\mu \nu } + ... + dp^\mu dp_\mu +   p^{\mu \nu } dp_{\mu \nu }
+ .....}~.
\eqno (2.15 ) $$

This action is the C-phase-space extension of the action of Nesterenko  and
involves quadratic derivatives in C-spaces which from the spacetime
perspective are effective {\em higher} derivatives theories [16]  where it
was shown why the scalar curvature in C-spaces is equivalent to a  higher
derivative gravity. One should expect a similar behaviour for the extrinsic
curvature of a polyparticle motion in C-spaces. This would be the C-space
version of the action for rigid particles [7]. Higher derivatives are the
hallmark of {\em W}-geometry (higher conformal spins).

Born-Infeld      s have been connected to  maximal-acceleration [8]. Such
models admits an straightforwad formulation using the geometric calculus of
Clifford algebras. In particular one can rewrite all the nonlinear equations
of motion in precise Clifford form [9]. This lead the author to propose the
$nonlinear$ extension of Dirac's equation for massless particles due to the
fact that spinors are nothing but right/left ideals of the Clifford algebra:
i.e., columns, for example, of the  Maxwell-Field strength bivector
$F = F_{\mu\nu} \gamma ^{\mu} \wedge \gamma ^{\nu}$.

Actions with higher derivatives may exhibit tachyonic behaviour and may
contain ghosts. In C-spaces the actions written for the Clifford polyvector
variables do not involve higher derivatives than the {\em quadratic} ones.
For this reason it is very compelling to suggest that
well-behaved physical theories in C-spaces may appear from the ordinary
spacetime perspective as tachyonic, since quadratic holographic derivatives
involving areas, volumes can be translated as higher derivatives in ordinary
spacetime variables [16]. Therefore, the question of whether or not tachyons
are unphysical depends on which space perspective one is taking. See [3] for
an interesting  discussion on this issue from the point of view of the many
worlds interpretation of QM.

To sum up,
we have now linked the maximal proper acceleration Relativity of the
spacetime tangent bundle with the minimal Planck scale Relativity in 
C-spaces
by a simple use of Clifford algebras in phase spaces . We obtained in one
stroke the action  of Nesterenko  compatible with the results of Brandt,
Schuller, Caianiello. Moreover, one could naturally generalize these actions
by working with the polyvector coordinates associated with a polyparticle  
in
a full-fledged Clifford-phase-space. In this fashion how one will have the
starting point for a phase space Relativity theory based on Clifford
algebras. Therefore, Extended Scale Relativity in C-spaces admits a natural
extension to phase spaces that will allow us to construct connections,
curvatures in C-phase spaces in the same way it was achieved for $curved$ C-
spaces. The scalar curvature in C-spaces was given by sums of products of
ordinary curvature with torsion [16]. The Einstein-Hilbert action in 
C-spaces
is a higher derivative gravity with torsion in the ordinary spacetime.  It 
is
unnecessary to insert these higher derivative terms by hand since they 
follow
from the geometry of C-spaces. This agrees with the low energy effective
string action obtained from non-linear $\sigma$      s. Recent results  to
construct a phase space Relativity appeared in [14], however these authors
are not invoking an extended Relativity principle based on a minimal Planck
scale nor a maximal proper acceleration and the use of Clifford algebras.

\bigskip

\centerline {\bf III. Why C-space Relativity over all the others }

\bigskip

There are several physical reasons why we believe that the Extended Scale
Relativity in C-spaces based on Clifford algebras is more physically
appealing than the construction of relativity  theories based  on kappa-
deformed Poincare algebras. The first reason is based on the fact that
quantum  group symmetries (algebras)  do
$not$ act on classical spacetimes. Whereas the C-space Lorentz $ group$  and
the Extended Relativity principle $already$ operates in a $classical$
Clifford space: i.e., the quantization procedure is $not$ responsible for 
the
Extended Relativity principle. One can then proceed to $quantize$ our
physical       using quantum deformation of Clifford algebras, in 
particular,
Braided Hopf Quantum Clifford algebras.

The Magueijo-Smolin      , a particular class of Doubly Special Relativity,
was shown recently to be unphysical [15]. This       corresponds to a
particular basis of kappa-deformed Poincare algebras. Whether these
unphysical results will also affect the outcome in other bases is not clear
yet. If so this raises questions about the validity of constructing
Relativity theories based on such quantum algebra.

Secondly, there is no reason why the Relativity theory based on kappa-
deformed Poincare algebras is more advantageous than the ones based on the
inhomogeneous Lorentz quantum groups  acting in  quantum-Minkowski 
spacetimes
[17].  Evenfurther, it was pointed out by Castellani that kappa-deformed
Poincare algebras are $not~ bicovariant$ which is a problem if one wishes to
construct physical theories, whereas the quantum algebras used in his $q$-
Gravity construction are bicovariant [17].

If the scale $ \Lambda  = 1/ \kappa$ was identified with the deformation
parameter one could equally as well relate the $q$-deformation parameter of
the inhomogeneous Lorentz quantum group to the minimum Planck scale. The
classical limit is recovered when $q \rightarrow 1 $ ; $ \Lambda \rightarrow 
0$.

Furthermore, there exists $multiparametric$ quantum deformations of 
classical
algebras and hence one could have assigned physical meanings (like the
minimal invariant scale)  to each single one of those quantum parameters.

We may notice that $q$ could be written in terms of an upper and lower scale 
as : $ q = e^{\Lambda/L } $. The classical limit $q = 1$, is obtained when $ 
\Lambda = 0 $ and also when $ L  = \infty$. This entails that there could be 
$two$ dual quantum gravitational theories with the same classical limit. 
Castellani has also pointed out that a sort of a  ``large/small"  duality 
exists in some of these      s based on those quantum algebras: there
is a symmetry $q \leftrightarrow 1/q $ reminiscent of the $T$-duality in
string theory.  One can see that by simply changing the signs of $ \Lambda$ 
or of $ L $ this is equivalent to replacing $ q $ for $ 1/q $. This 
large/small duality is just another manifestation of the
ultraviolet/infrared entanglement of QFT defined over Noncommutative spaces
(geometries). Nottale postulated that if there is a minimum Planck scale, by
duality, there should be an upper impassible invariant scale $ L$ in Nature. 
This
was his proposal for the resolution of the cosmological constant problem 
[2].

These arguments raises the possibility that the Double Special Relativity
theories based on the kappa-deformed Poincare algebra [10-13] might be
obtained by a Moyal deformation quantization procedure in phase spaces where
the Moyal deformation parameter is precisely  $\Lambda = 1/ \kappa$.

Vasiliev [18] has constructed a consistent infinite-component  higher spin
field theory  in Anti de Sitter spaces based on similar star products whose
deformation parameter is the inverse size of the  Anti de Sitter $throat$.
The ``classical" flat spacetime  limit is recovered when $l \rightarrow
\infty$ so $1/l  \rightarrow 0$. This is an identical situation in
constructing the Poincare algebra from the de Sitter algebra by a Wigner-
Inon\"u contraction, when $l = \infty$ the commutators

 $$ [ P^\mu, P^\nu ] = { 1 \over  l^2 } [ J^{ 5 \mu } , J^{5\nu } ] = 0 .
\eqno(3.1) $$

It is true that Hopf quantum algebras are richer than a naive Moyal
deformation of ordinary
algebras. In particular, in order to write down the phase space algebra
consistent with the Hopf quantum algebra of spacetime it is necessary to use
the Heisenberg-double prescription based on the interplay between the
algebraic sector of the Hopf algebra and the $co-algebraic$ sector [10].

Put it simply, one cannot naively use the phase space algebra $[x, p]$ in
order to generate  the kappa-deformed Poincare algebra in the bicrossproduct
basis, or the Snyder basis, for example, by naïvely writing the boosts-
momentum commutator:
 $$ [ N_j , p_k ]   =  [ x_0 p_j - x_j p_0 , p_k ] =  [ x_0, p_k ] p_j   -
[ x_j, p_k ] p_0 ~,
\eqno ( 3.2 ) $$
due to the fact that the $deformed$ boost generator has no longer the same
form as in the classical case:

 $$ N^{ ( \kappa) }_j  =  N_j  + O ( 1 / \kappa ) = x_0 p_j - x_j p_0 + O (
1/ \kappa ) .
\eqno(3.3) $$

 This is the reason why one must use the Heisenberg-double prescription in
order to extract the phase space commutator algebra from the kappa-deformed
boosts-momentum commutators and the remaining ones.

 Nevertheless, one alternative is not to $separate$ the configuration space
from the momentum space but could be by starting to work $directly$ in the
phase space and writing down the Moyal star product:

$$ X ( q, p , \kappa ) * P ( q, p, \kappa ) =$$  $$\sum_0^\infty    { ( 1/
\kappa)^n  \over n!}
\omega^{ A_1 B_1 } \omega^{ A_2 B_2 } \omega^{ A_n B_n }
( \partial_{ A_1 } \partial_{A_2}..\partial_{A_n} X ( q, p, \kappa ) ) (
\partial_{B_1}
\partial_{B_2} …..\partial_{B_n} P ( q, p, \kappa) ) ~, \eqno(3.4 ) $$
using the $ 1/ \kappa$ as the $deformation$ parameter in order to evaluate
the Moyal brackets. The nondegenerate invertible symplectic form 
$\omega_{AB}
$, associated with the $ 2n$-dimensional phase space is what defines the
standard Poisson brackets:

$$y^A \equiv   q^0, q^1, q^2,...,q^n ; p^0, p^1, p^2 ,...,p^n )  
\Rightarrow
\{ y^A, y^B \}_{PB} = \omega^{AB} . \eqno ( 3.5) $$

The Moyal bracket with respecto to the $ ( q, p ) $ variables is defined in
terms of the star product as:

$$\{   X , P \}_{ MB } \equiv   { X * P - P* X \over ( 1/ \kappa) }
\leftrightarrow
{ 1 \over i } [ X, P ]  . \eqno ( 3.6 )   $$

Therefore, the following Moyal bracket must obey:

$$ \{  N_i , P_j \}_{MB}     =  \{  X_0 P_i - X_i P_0 , P_j  \}_{MB}  =
F_{ij} [ q ( X, P, \kappa ) , p ( X, P , \kappa ) , \kappa ]  =  $$
$$ \delta_{ij} [ {\kappa \over 2 } ( 1 - e^{ - 2 P_0 / \kappa } )  + { 1
\over 2 \kappa} {\vec P}^2 ]  - { 1 \over \kappa} P_i  P_j  ~, \eqno (3.7)$$
after re-expressing the bracket back into the $original$ $ X, P$ variables,
in order for the Moyal bracket to be $ isomorphic$  to the commutator
$ [ {\hat N}_j, {\hat P}_k ]_{ \kappa} $ of the kappa-deformed Poincare
algebra
in the bicrossproduct basis, for example. For this to occur one must find 
if,
and only if, there is a one-to-one and $invertible$ analytic map obeying 
such
conditions:

$$ X = X ( q, p , \kappa ) = \sum X_{ ( n ) }   q, p ) ( 1/ \kappa)^n~,$$
$$P =   ( q, p, \kappa ) = \sum P_{ ( n) } ( q, p ) ( 1 / \kappa)^n \eqno
(3. $$
and thus $\Leftrightarrow q = q ( X, P, \kappa ) ,
~~ p = p ( X, P, \kappa )$,  
with the provision that
$$ X^\mu _{ (o)}  = q^\mu , ~~~ P^\mu _{ ( o ) } = p^\mu . \eqno ( 3.9 ) $$

 It is fairly clear that the $classical$ basis of kappa-deformed Poincare 
is
given precisely by
$X = q, P = p$ so the Moyal bracket algebra collapses to the ordinary 
Poisson
bracket algebra and one recovers the $undeformed$ classical Poincare 
algebra.

It is essential to notice that the map (3. is $noncanonical $, i.e., it
does $not$ preseve the symplectic  form !  For this reason, the quantum
algebra is not $unique$ in so far that the commutators $change$ under a
$noncanonical $ change of basis; i.e., the quantum algebra is $not$ 
invariant
under a $noncanonical $ basis change of coordinates.

If one wishes to reproduce the kappa-deformed Poincare algebra in the Snyder
basis, for example, from the classical algebra, one will require a
$different$ phase space transformation than before
$$ X' = X' ( q, p , \kappa ) , ~~~ P' =  P' ( q, p, \kappa )~, \eqno (3.10)
$$
to ensure that $\{  N_{j }, P_{k} \}_{MB}$ is isomorphic to the commutator
$[{\hat  N}_j,  {\hat P}_k]$ in the Snyder basis. Of course, one still has 
to
check that the undeformed sector of the quantum algebra remains $intact$. We
don't know if this procedure will reproduce the kappa-deformed Poincare
algebra in any basis. Perhaps this is an empty academic exercise because one
must be forced to always use the Heisenberg-double prescription, which is
deeply ingrained in the algebra and co-algebra sector of the Hopf algebra, 
in
order to read-off the phase space algebra and which does not have a naïve
Moyal star product interpretation.

In four dimensions we have the choice of eight functions $ X (q, p,
\kappa ) , P (q, p, \kappa )$ to match the kappa-deformed Poincare algebra 
in
a given basis. We must see whether or not one can find indeed eight 
functions
that reproduce all the $independent$ commutators of the quantum algebra. Not
all commutators are independent due to the constraints imposed by the Jacobi
identities. This is not an easy task.

The reason we raise this possibility is because it is plausible that the 
role
of the Planck scale
$ \Lambda = 1/ \kappa$ in kappa-deformed Poincare Relativity might be
identical to the $deformation$ parameter  of the  noncommutative Moyal star
product construction in phase spaces. This is in sharp distinction to the
natural role of the Planck scale in C-space Relativity: it must be there on
pure dimensional grounds to combine objects
(strings, branes) of different dimensionalities and  consistent with the
postulated minimum scale principle. To have real valued intervals  $dX. dX  
 >
0$ in C-space requires that the variable scales which encode the magnitudes
of the generalized (holographic)  velocities cannot be smaller than 
$\Lambda$
[1].

One main advantage of using Clifford algebras versus the kappa-deformed
Poincare algebras is that the C-space Lorentz transformations form a $ group
$ in a very natural fashion! Recently it has been argued [13] that the kappa 
-
deformed algebra forms a group in the bicrossproduct basis but unfortunately
there was a caveat because the infinite series expansion obtained from the
Baker-Cambell-Hausdorff formulae might $not$ converge. Even if this were the
case, it is not true that one has the required group structure in the other
infinite bases which is very unphysical since there is no reason why one
basis should be more ``physical" than the others, i.e., it questions the
minimal length Relativity principle based on kappa-deformed Poincare
algebras.

In addition, there still remains the serious problem of how to add momenta.
The momenta-addition law is $nonabelian$ in kappa-deformed Poincare due to
the nontrivial nature of the co-product and poses problems for the physics 
of
many-particle systems.The nonabelian addition law contradicts well known
experimental facts. In C-space, polyvectors are added in a linear and 
abelian
fashion. Finally, there is no need to add an extra discrete dimension to
explain the Snyder noncommutative algebra for the spacetime coordinates [1]
nor to work in six-dimensions in order to derive the algebra of the 
conformal
group  $SO(4,2)$. All can be explained naturally from the Clifford algebra 
of
the four-dimensional spacetime [16]. These results are also shared by the
pseudo-complexified Minkowski spacetime [8] approach to the maximal
acceleration.

\bigskip

\centerline{\bf  Acknowledgements}

We are kindly indebted to H. Rosu and L. Perelman for their assistance in
preparing the manuscript.

\bigskip

\centerline { \bf References }

\bigskip

\noindent
1 - C. Castro, ``The programs of the Extended Relativity in C-spaces, 
towards
the physical foundations of String theory", hep-th/0205065. To appear in the
proceedings of the NATO advanced workshop on the nature of time, geometry,
physics and perception. Tatranska Lomnica, Slovakia, May 2002.  Kluwer
Academic Publishers;
``Noncommutative Quantum Mechanics and Geometry from the quantization of C-
spaces", hep-th/0206181.

\bigskip

\noindent
2 - L. Nottale, ``Fractal Spacetime and Microphysics, towards Scale
Relativity", World Scientific, Singapore, 1992;
``La Relativite dans tous ses etats", Hachette Literature Paris, 1999.

 \bigskip

\noindent
3 - M. Pavsic, ``The landscape of Theoretical Physics: A global view  from
point particles to the brane world and beyond, in search of a unifying
principle",  Kluwer Academic Publishers 119, 2001.

\bigskip

\noindent
4 - E. Caianiello, ``Is there a maximal acceleration?", Lett. Nuovo Cimento
{\bf 32} (1981) 65.

\bigskip

\noindent
5 - V. Nesterenko, Class. Quant. Grav. {\bf 9} (1992) 1101; Phys. Lett. {\bf
B 327} (1994) 50;
V. Nesterenko,  A. Feoli, G. Scarpetta, ``Dynamics of relativistic particles
with Lagrangians dependent on acceleration", hep-th/9505064.

\bigskip

\noindent
6 - H. Brandt,  Chaos, Solitons and Fractals {\bf 10} (2-3) (1999) 267;
Found. Phys. Lett. {\bf 2 } (1) (1989) and {\bf 4} (6) (1991).

 \bigskip

\noindent
7 - M. Pavsic, Phys. Lett. {\bf B 205} (198 231; Phys. Lett. {\bf B 221}
(1989) 264;
H. Arodz, A. Sitarz, P. Wegrzyn, Acta Physica Polonica {\bf B 20} (1989) 
921.

\bigskip

\noindent
8 - F. Schuller, ``Born-Infeld Kinematics and corrections to the Thomas
precession",
hep- th/0207047; Ann. Phys. {\bf 299} (2002) 174, hep-th/0203079.

 \bigskip

\noindent
9 - A. Chernitskii, ``Born-Infeld electrodynamics, Clifford numbers and
spinor representations", hep-th/0009121.

\bigskip

\noindent
10 - J. Lukierski, H. Ruegg, W. Zakrzewski, Ann. Phys. {\bf 243 } (1995) 90;
J. Lukierski , A. Nowicki, H. Ruegg, V. Tolstoy, Phys. Lett. {\bf B 264}
(1991) 331;
J. Lukierski, A. Nowicki : Proceedings of Quantum Group Symposium at Group 
21
(July 1996, Goslar), Eds. H. D. Doebner and V.K. Dobrev , Heron Press, 
Sofia,
1997, p 186, math.QA/9803064, hep-th/9812063;
J. Lukierski, A. Nowicki, ``Doubly Special Relativity versus kappa-deformed
dynamics", hep-th/0203065.

 \bigskip

\noindent
11 - S. Majid, H. Ruegg, Phys. Lett. {\bf B 334} (1994) 348.

 \bigskip

\noindent
12 - G. Amelino-Camelia, Int. J. Mod. Phys. {\bf D 11} (2002) 35, gr-
qc/0012051;
Phys. Lett. {\bf B 510} (2001) 255;
J. Kowalski-Glikman, S. Nowak, ``Doubly special relativity theories as
different bases of kappa-Poincare", hep-th/0203040.

 \bigskip

\noindent
13 - N. Bruno, ``Group of boosts and rotations in  Double Special
Relativity", gr-qc/0207076.

 \bigskip

\noindent
14 - G. Watson, J. Klauder, ``Metric and curvature in gravitational phase
space", gr-qc/0112053.

 \bigskip

\noindent
15 - J. Rembielinski, K. Smolinski, ``Unphysical predictions of some Doubly
Special Relativity theories", hep-th/0207031.

 \bigskip

\noindent
16 - C. Castro, M. Pavsic, ``Higher derivative gravity and torsion from the
geometry of C-spaces", Phys. Lett. {\bf B 539} (2002) 133 , hep-th/0110079;
``The Clifford algebra of spacetime and the conformal group", 
hep-th/0203194.

\bigskip

\noindent
17 - L. Castellani, ``The Lagrangian of q-Poincare Gravity", Phys. Lett. 
{\bf
B 327} (1994) 22, hep-th/940233;
``Differential calculus on Iso$_q$(N), quantum Poincare algebra and q-
gravity", Commun. Math. Phys. {\bf 171} (1995) 383;
P. Aschieri, L. Castellani , A. Scarfone, ``Quantum orthogonal planes,
bicovariant calculus and differential geometry on quantum Minkowski space",
Eur. Phys. J. {\bf C7} (1999) 159, q-alg/9709032;
A. Dimakis, F. Muller-Hoissen, ``Discrete differential geometry", J. Math.
Phys. {\bf 40} (1999) 1518.

 \bigskip

\noindent
18 - M. Vasiliev, ``Progress in higher spin theories", hep-th/0104246.

\end{document}